\documentclass[aps,prx,twocolumn,superscriptaddress,longbibliography,notitlepage]{revtex4-2}

\usepackage[T1]{fontenc}
\usepackage[utf8]{inputenc}
\usepackage[english]{babel}
\usepackage{amsmath}
\usepackage{amsfonts}
\usepackage{amssymb}
\usepackage{graphicx}
\usepackage{hyperref}
\usepackage{physics}
\usepackage{pgf}
\usepackage{braket}
\usepackage{textcomp}
\usepackage{wasysym}
\usepackage{soul, color}

\graphicspath{{./Figures/}}

\DeclareUnicodeCharacter{2212}{$-$}
\let\pgfimageWithoutPath\pgfimage
\renewcommand{\pgfimage}[2][]{\pgfimageWithoutPath[#1]{Figures/#2}}

\newcommand{\Tone}{$T_1$}
\newcommand{\Gone}{$\Gamma_1$}
\newcommand{\Ttwoecho}{$T_2^\textrm{e}$}
\newcommand{\Gtwo}{$\Gamma_2^\textrm{e}$}
\newcommand{\Gtwophi}{$\Gamma_{2,\Phi}^{\textrm{e}}$}
\newcommand{\Af}{$\sqrt{A_{\Phi,1/f}}$}
\newcommand{\Sw}{$\sqrt{S_{\Phi,\textrm{BB}}}$}
\newcommand{\onef}{$1/f$}
\newcommand{\NH}{NH$_3$}
\newcommand{\sfsix}{SF$_6$}

\begin{document}

\title{Effects of surface treatments on flux tunable transmon qubits}

\author{M. Mergenthaler}
\email{mme@zurich.ibm.com}
\affiliation{IBM Quantum, IBM Research - Zurich, S\"aumerstrasse 4, 8803 R\"uschlikon, Switzerland}
\author{C. M\"uller}
\affiliation{IBM Quantum, IBM Research - Zurich, S\"aumerstrasse 4, 8803 R\"uschlikon, Switzerland}
\author{M. Ganzhorn}
\affiliation{IBM Quantum, IBM Research - Zurich, S\"aumerstrasse 4, 8803 R\"uschlikon, Switzerland}
\author{S. Paredes}
\affiliation{IBM Quantum, IBM Research - Zurich, S\"aumerstrasse 4, 8803 R\"uschlikon, Switzerland}
\author{P. M\"uller}
\affiliation{IBM Quantum, IBM Research - Zurich, S\"aumerstrasse 4, 8803 R\"uschlikon, Switzerland}
\author{G. Salis}
\affiliation{IBM Quantum, IBM Research - Zurich, S\"aumerstrasse 4, 8803 R\"uschlikon, Switzerland}
\author{V.P. Adiga}
\affiliation{IBM Quantum, IBM T.J. Watson Research Center, Yorktown Heights, NY 10598, USA}
\author{M. Brink}
\affiliation{IBM Quantum, IBM T.J. Watson Research Center, Yorktown Heights, NY 10598, USA}
\author{M. Sandberg}
\affiliation{IBM Quantum, IBM T.J. Watson Research Center, Yorktown Heights, NY 10598, USA}
\author{J.B. Hertzberg}
\affiliation{IBM Quantum, IBM T.J. Watson Research Center, Yorktown Heights, NY 10598, USA}
\author{S. Filipp}
%\email{Current affiliation: Department of Physics, Technical University of Munich, 85748 Garching, Germany}
\affiliation{IBM Quantum, IBM Research - Zurich, S\"aumerstrasse 4, 8803 R\"uschlikon, Switzerland}
\author{A. Fuhrer}
\affiliation{IBM Quantum, IBM Research - Zurich, S\"aumerstrasse 4, 8803 R\"uschlikon, Switzerland}

\date{\today}

\begin{abstract}
	One of the main limitations in state-of-the art solid-state quantum processors are qubit decoherence and relaxation due to noise in their local environment.
	For the field to advance towards full fault-tolerant quantum computing, a better understanding of the underlying microscopic noise sources is therefore needed.
	Adsorbates on surfaces, impurities at interfaces and material defects have been identified as sources of noise and dissipation in solid-state quantum devices.
	Here, we use an ultra-high vacuum package to study the impact of vacuum loading, UV-light exposure and ion irradiation treatments on coherence and slow parameter fluctuations of flux tunable superconducting transmon qubits.
	We analyse the effects of each of these surface treatments by comparing averages over many individual qubits and measurements before and after treatment.
	The treatments studied do not significantly impact the relaxation rate \Gone ~and  the echo dephasing rate \Gtwo, except for Ne ion bombardment which reduces \Gone.
	In contrast, flux noise parameters are improved by removing magnetic adsorbates from the chip surfaces with UV-light and \NH ~treatments.
	Additionally, we demonstrate that \sfsix ~ion bombardment can be used to adjust qubit frequencies in-situ and post fabrication without affecting qubit coherence at the sweet spot.
\end{abstract}
\maketitle

\section{Introduction}

%Quantum technology has been rapidly developing this past decade, with publicly available superconducting quantum processors accessible through the cloud~\cite{IBMQ}, experimental demonstrations of ever more complex quantum algorithms~\cite{Arute2019, 20JurcevicAA, 20EganAA} and many other steps towards the ultimate goal of a universal, fault-tolerant quantum computer.
Current solid-state quantum processors are still limited by intrinsic error mechanisms that lead to relaxation of the qubit state, the loss of phase coherence over time, i.e. dephasing, as well as readout and unitary errors when performing gate operations.
Coherence times in superconducting circuits have already improved by about five orders of magnitude over the last two decades~\cite{Nakamura1999, 13OliverAA} and many groups around the world now routinely report qubit lifetimes in the 10-100~$\mu$s range, with average dephasing times that depend on the chosen qubit design.
Fixed frequency qubits (FFQs), where the level splitting is set by design during fabrication,
show the best coherence of often several hundred microseconds.
However, they require higher fabrication precision as well as more elaborate qubit coupling and quantum gate schemes~\cite{20GanzhornAA}.%\af{what other citations}
Conversely, frequency tunable qubits (FTQs) that either include a magnetic field tunable SQUID~\cite{Chen2021, 20AndersenAA, Klimov2018, Versluis2017} or an electric field tunable nanowire junction~\cite{18CasparisAA}, typically have shorter coherence times of around ten microseconds due to the additional sensitivity to environmental fluctuations.

In both cases decoherence is caused by the coupling of the quantum circuit to uncontrolled, environmental degrees of freedom.
Here, one distinguishes two basic effects: energy exchange with the environment which is characterized by the relaxation time $T_{1}$, as well as pure dephasing, where the qubit level splitting $f_{01}$ is affected by environmental fluctuations that lead to the decay of the phase of superposition states on a timescale $T_{\varphi}$ in the ensemble average over many repeated experiments.

\begin{figure}[t]
	\centering
	\includegraphics[width=0.97\columnwidth]{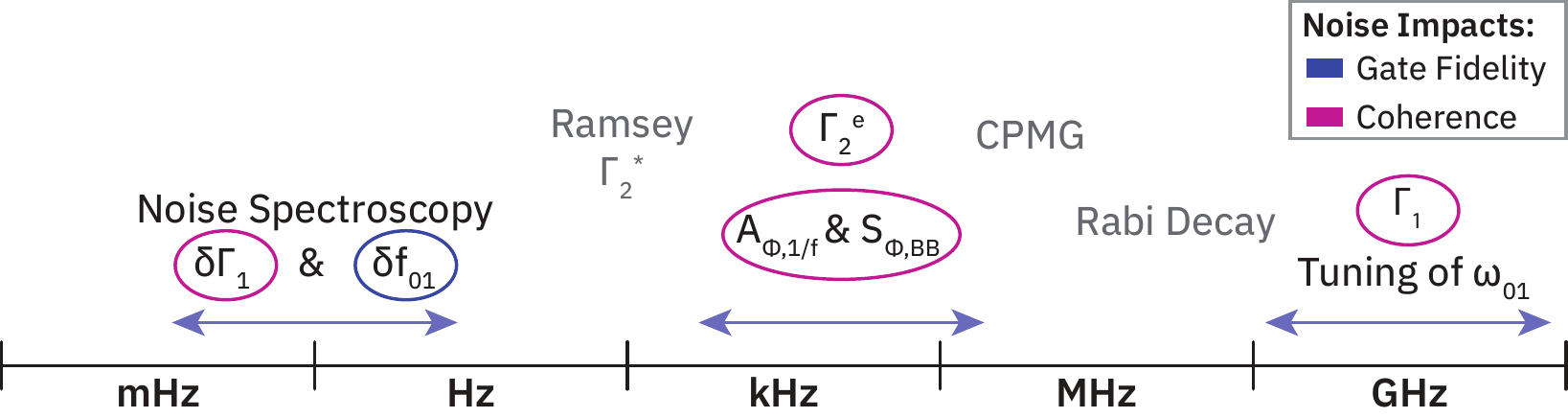}
	\caption{Spectral sensitivity of qubit noise measurements: $\delta\Gamma_1$ and $\delta f_{01}$ denote low-frequency parameter fluctuations as measured by the widths of long-term $\Gamma_1$ and $f_{01}$ histograms. $\Gamma_2^*$ probes frequency fluctuations over the full Ramsey averaging time whereas Hahn, and CPMG echo sequences probe  noise only above a cut-off frequency. $A_{\Phi,1/f}$ and $S_{\phi, BB}$ denote $1/f$ and broadband flux-noise powers extracted from the flux dependence of \Gtwo. Rabi decay probes noise at the qubit Rabi frequency and $\Gamma_1$ spectroscopy is sensitive to noise at the qubit frequency. }
	\label{fig:fig1}
\end{figure}

There are two main sources for environmental perturbations: fluctuating magnetic and electric fields.
The former is thought to come from adsorbed fluctuating spins on surfaces or at material interfaces of quantum circuits.
Both adsorbed oxygen molecules~\cite{Kumar2016, Wang2015a} and atomic hydrogen~\cite{18GraafAA,17QuintanaAA} have been indicated as possible origins for such fluctuations.
These mainly impact FTQs that employ SQUIDs to modulate the Josephson energy by a magnetic flux, and hence lead to fluctuations of the qubit level splitting.
Similarly, external magnetic field noise and instrumentation noise can also cause such fluctuations. Electric field noise on the other hand typically originates from microscopic fluctuations of the dielectric properties of the insulators used in circuit fabrication, fluctuations in the configuration of surrounding charge traps, as well as noise in the control lines~\cite{Dunsworth2017,16GambettaAA,16ChiaroAA,15WangAB,15DialAA,15BrunoAA,14QuintanaAA,13BarendsAB,12GeerlingsAA,11WennerAA,11SageAA,11PappasAA,10WisbeyAA,09WangAA,08HouckAA,08GaoAB}.
Due to the prevalence of electric field noise, transmon qubits have been engineered to make the qubit splitting insensitive to such fluctuations, restricting their impact to transverse coupling and thus mostly qubit energy decay and excitation.
However, qubit frequency fluctuations may also occur trough variations in dispersive shifts of near resonant defect states that interact with their respective environments~\cite{Muller2019, 15FaoroAA, Muller2015}.
Dielectric fluctuators are most important in regions where the fields of the relevant qubit modes are strongest i.e. in the junction oxide, at surfaces near the junction and near edges of the metallic qubit elements~\cite{Schlor2019, 20BilmesAA}.
Junction oxides and subsurface interfaces can in principle be improved by optimizing fabrication processes. However, the surfaces that are exposed to ambient after qubit fabrication are especially hard to control and both native surface oxides and adsorbed molecules can cause qubit decoherence.

An additional source of noise in superconducting qubits can arise from quasiparticles, which affect the qubit level splitting through parity effects~\cite{Catelani2011,19SerniakAA} as well as absorb energy when tunneling across the junctions~\cite{12CatelaniAA, 14WangAA, 16GustavssonAA}.

So far we have differentiated environmental noise sources by their physical coupling to the qubit.
Another distinction between environmental fluctuations can be made with respect to their frequency. Here, low frequency noise, at time scales much slower than the internal dynamics of the circuits, will lead to dephasing and calibration errors.
Typical low-frequency noise spectra of both electric as well as magnetic field fluctuations show a characteristic $1/f$ behavior with a divergence at low frequencies.
Conversely, high-frequency noise at the qubit energy can lead to relaxation and excitation, where the environment absorbs or excites a quantum of energy from the quantum circuit. This is mostly relevant for electrical noise and quasiparticles, as magnetic flux noise spectra have been shown to behave as $\sim 1/f$ up to very high frequencies~\cite{13YanAA, 12SankAA, 17QuintanaAA}.

It is clear that different coherence parameters probe different parts of the environmental noise spectrum and are sensitive to either magnetic or electrical noise or both.
Fig.~\ref{fig:fig1} schematically shows a frequency spectrum probed by specific coherence measurements.
The relaxation rate $\Gamma_1$ of a qubit is sensitive to the noise spectrum at the qubit frequency which in the case of FTQs can be adjusted to perform spectroscopy of strongly coupled resonant two-level systems (TLS)~\cite{05MartinisAA, 20BilmesAA}.
Rabi decay measurements are additionally sensitive to noise at the Rabi frequency of the qubit which is typically in the 100~MHz range~\cite{05IthierAA,13YanAA}.
The middle frequency range (cf. Fig.~\ref{fig:fig1}) can be accessed through Ramsey type experiments where $\Gamma_2^*$  is sensitive to fluctuations over the entire  measurement time whereas Hahn, and CPMG echo sequences filter out the low frequency noise components~\cite{11BylanderAA}.
In FTQs the dependence of \Gtwo~ on flux allows extraction of separate $1/f$ and a broadband flux-noise contributions in the frequency band where noise affects \Gtwo~\cite{Luthi2018}.
At the low frequency end, slow fluctuations in $\Gamma_1$ and the qubit frequency $f_{01}$ are important for qubit calibrations, gate fidelities and the reproducibility of quantum circuits in general.

In the present paper, we use the coherence parameters circled in Fig.~\ref{fig:fig1} in order to probe the impact of UV light, ion milling  (with ions from Ne and SF$_6$ gas) and passivation with NH$_3$ on high coherence FTQs.
To preserve the effects of the treatments during loading of the chips into the dilution refrigerator, we use a UHV package~\cite{Mergenthaler2021} that maintains vacuum around two qubit chips with up to eight qubits until the chips are cold.
UV light and ion milling can be used to remove unwanted surface contamination and in the case of ion milling even to remove thin surface oxide layers.
However, these treatments create unsaturated bonds and may leave voids or implanted ions in the exposed surfaces.
For this reason we include a subsequent passivation treatment with NH$_3$ in our study.
Our results not only show the different effects of the treatments on the outlined coherence parameters but also indicate that ion milling can be used to trimm the qubit frequency of FFQs after fabrication without a significant reduction in coherence.

The paper is organized in seven sections. Following the introduction we describe each of the surface treatment procedures in Section~\ref{treatment-descriptions} and specify the coherence parameters (Section~\ref{coherence-parameters}) that we use to directly compare the impact of different treatments in Section~\ref{treatment-comparisons}. In Section~\ref{noise-spectroscopy} we focus on the effect of these surface treatments on slow parameter fluctuations that are specifically relevant for maintaining qubit calibration in quantum computing systems. In Section~\ref{frequency-tuning} we discuss the impact of ion-bombardment treatments on average qubit frequency and qubit relaxation and we summarise our findings in Section~\ref{conclusion}.

\section{Treatment Descriptions}
\label{treatment-descriptions}

We study a set of eight surface treatments by directly comparing qubit chips in subsequent cool-downs with and without the specific treatment.
The treatment system and the UHV package are described in detail in Ref.~\cite{Mergenthaler2021}.
A list of all the treatments in each pair of subsequent cool-downs is given in Table~\ref{tab:tab1}, where the first column gives the numbering and label for each comparison as used below.
The second column highlights the investigated treatment, i.e. the difference between the two cool-downs, in bold face.

 \begin{table}[h]
	\centering
	\begin{tabular}{l|rcl}
		Comparison & Comparison &of& treatments\\
		label & (pre &/& post)\\
		\hline\hline
		\emph{(i) UHV} &  ambient &/& {\bf UHV}  \\
		\emph{(ii) UV}  & UHV &/& {\bf UV} + UHV  \\
		\emph{(iii) UV + NH$_3$} & NH$_3$ &/& {\bf UV} + NH$_3$ \\
		\emph{(iv) R} & UV + UHV &/& UV + UHV\\
		\emph{(v) Ne} & UV + UHV &/& UV + {\bf Ne} + UHV\\
		\emph{(vi) Ne + NH$_3$} & UV + UHV &/& UV +{ \bf Ne + NH$_3$}\\
		\emph{(vii) SF$_6$} & UV + UHV &/& UV + {\bf SF$_6$} + UHV\\
		\emph{(viii) SF$_6$ / SF$_6$} & UV + SF$_6$ + UHV &/&  UV + SF$_6$ + UHV\\
		\hline
	\end{tabular}
	\caption{List of treatment comparisons: The difference between reference (pre) and modified (post) treatment sequence in each comparison is highlighted in bold.}
	\label{tab:tab1}
\end{table}

\setlength{\parindent}{0pt}
Investigated treatments are:
\paragraph*{\textbf{\emph{ambient}}} Standard packaging of the qubit chips under ambient conditions and at room temperature. Here, the sample environment is only pumped when the cryostat is cooled. This is the only treatment where the qubit chip is not annealed at T$_A= 80^\circ$C for six or more hours under vacuum before applying other treatments and closing the package.

\paragraph*{\textbf{\emph{UHV}}} Closure of the UHV package at p$_{base} < 10^{-9}$~mbar and with active pumping inside the package using a titanium getter layer during transfer to the cryostat.
From separate measurements we can put an upper limit of p$_{tr} < 5\times10^{-8}$~mbar on the pressure during transfer~\cite{Mergenthaler2021}.

\paragraph*{\textbf{\emph{UV}}} Chips are exposed to UV light (peak wavelength $\lambda = 140~$nm) for 10~mins at p$_{base} < 10^{-9}$~mbar to desorb water and other molecules from the sample surface. Subsequently, without breaking vacuum, the sample is packaged under UHV conditions.

\paragraph*{\textbf{\emph{NH$_3$}}} Back-fill with NH$_3$ to p$_{tr} = 5\times10^{-3}$~mbar after UV exposure and before closing the package.

\vspace{4pt}
The above treatments do not affect amorphous surface oxides or processing residues on the qubit chips.
To investigate the influence of such layers we use an ion gun with two different gas sources to sputter-clean the surface under a shallow angle of 30$^{\circ}$ from the chip surface.
 \vspace{4pt}

\paragraph*{\textbf{\emph{Ne}}} Ion milling of the chips for 20~minutes with neon ions at an energy E$_{Ne} = 0.75$~keV and p$_{Ne}$ = $10^{-4}$~mbar. This occurs after pump-down and UV light irradiation. The package is pumped down to p$_{base} < 10^{-9}$~mbar before closing and transfer.

\paragraph*{\textbf{\emph{Ne+NH$_3$}}} Back-fill with NH$_3$ to p$_{tr} = 5\times10^{-3}$~mbar after neon treatment with increased ion energy and twice the treatment duration (E$_{Ne} = 1.25$~keV and t = 40~min).

\paragraph*{\textbf{\emph{SF$_6$}}} Denotes a 20 minute ion milling treatment with SF$_6$ ion species at an energy E$_{SF_6} = 1.25$~keV and with a SF$_6$ pressure of p$_{SF_6} = 10^{-4}$~mbar.
After the ion mill, the package is pumped down to p$_{base} < 10^{-9}$~mbar before closing and transfer.

\setlength{\parindent}{15pt}

\section{Coherence Parameters}
\label{coherence-parameters}

\begin{figure}[t]
	\centering
	\includegraphics[width=0.97\columnwidth]{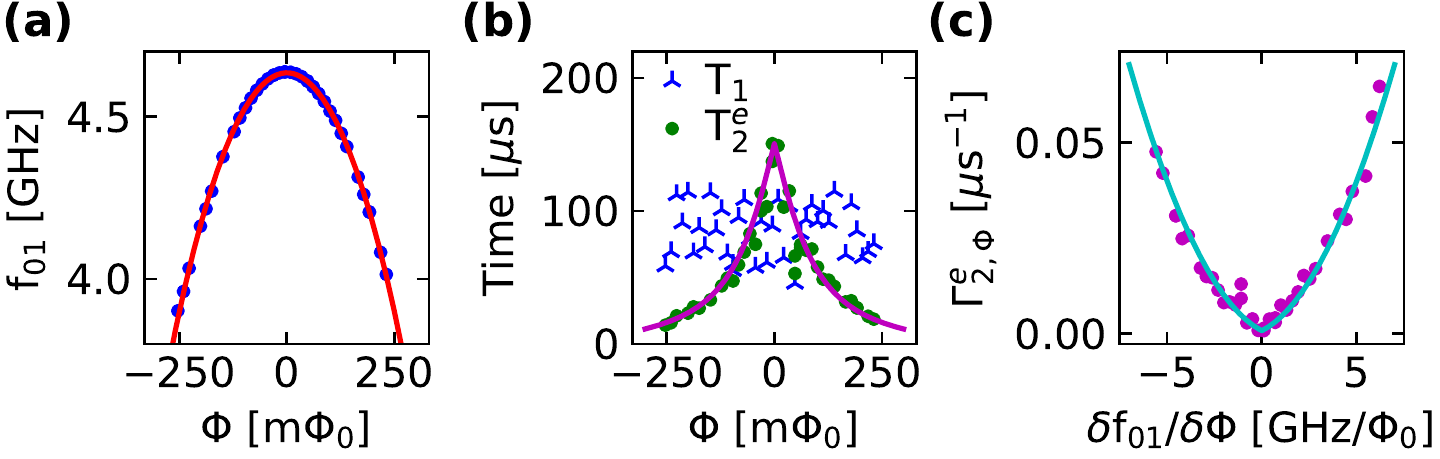}
	\caption{Flux ($\Phi$) dependent parameter analysis.
	(a) Qubit frequency as a function of $\Phi$.
	(b) \Tone~ and \Ttwoecho~ recorded as a function of flux.
	(c) Echo dephasing rate \Gtwophi~as a function of the flux sensitivity.}
	\label{fig:fig2}
\end{figure}

In order to compare the impact of surface treatments (c.f. Tab.~\ref{tab:tab1}) on qubit coherence, we perform flux-dependent measurements of the qubit frequency f$_{01}$, lifetime \Tone~and echo decay time \Ttwoecho~as shown in Fig.~\ref{fig:fig2}~(a-b).

Across all qubits and treatments \Tone~is usually independent of flux but typically shows strong fluctuations~\cite{Muller2015, Klimov2018, Schlor2019, Burnett2019b}, as indicated by the blue crosses in Fig.~\ref{fig:fig2}~(b) for a specific qubit.
In contrast, \Ttwoecho~increases symmetrically towards the sweet-spot (SSP) to almost $1.5\times$ the \Tone-average, whilst not exhibiting the same fluctuations as \Tone, see Fig.~\ref{fig:fig2}~(b) (green dots).
This behavior is expected since the impact of flux-noise increases further away from the SSP for FTQs and thus reduces \Ttwoecho.

To quantify the impact of flux-noise, we plot the echo dephasing rate \Gtwophi~$=\frac{1}{T_2^{\textrm{e}}}-\frac{1}{2T_1}$ as a function of the flux-sensitivity $D_{\Phi}=|\delta f_{01}/\delta\Phi|$ (see Fig.~\ref{fig:fig2}~(c)).
We then fit the data with a quadratic polynomial $\Gamma_{2,\Phi}^\textrm{e}=aD_\Phi^2+bD_\Phi+c$, where $a$ is a measure for non-diverging, broadband (BB) noise, $b$ for $1/f$ flux noise and $c$ is a constant offset from other, non-flux related dephasing contributions.

The amplitude $A_{\Phi,1/f}$ of the $1/f$ component of the flux-noise power spectral density (PSD) $S_{\Phi,\textrm{1/f}}$ is obtained from~\cite{Martinis2003, Yoshihara2006, Hutchings2017, Luthi2018}
\begin{equation}
	A_{\Phi,1/f}=\left(\frac{b}{2\Phi\sqrt{ln(2)}}\right)^2 \,,\:S_{\Phi,\textrm{1/f}}=\frac{A_{\Phi,1/f}}{f} \,,
	\label{eq:A}
\end{equation}
and similarly the magnitude of the broadband component of the flux-noise PSD is
\begin{equation}
	S_{\Phi,\textrm{BB}}=\frac{a}{\pi^2} \,.
	\label{fig:Sw}
\end{equation}
It is important to mention that \Ttwoecho~measurements only probe the noise spectrum in an intermediate frequency range (see Fig.~\ref{fig:fig1}) given by the filter function of the Hahn echo sequence~\cite{11BylanderAA}.
The extracted magnitudes of the flux-noise parameters may therefore not give the full picture but are still useful for comparing treatments.
A different approach is to fit the PSD of the qubit frequency measured with fast Ramsey spectroscopy over long periods of time.
This measures a mix of all low-frequency fluctuations that affect the qubit frequency.
We found that it is challenging to acquire long enough datasets for enough qubits to be able to make a reliable comparison between treatments.
Instead we have focused on other low-frequency coherence parameters as described later in the text (cf. Section~\ref{noise-spectroscopy}).

\section{Treatment Comparisons}
\label{treatment-comparisons}

\begin{figure}[t]
	\centering
	\includegraphics[width=\columnwidth]{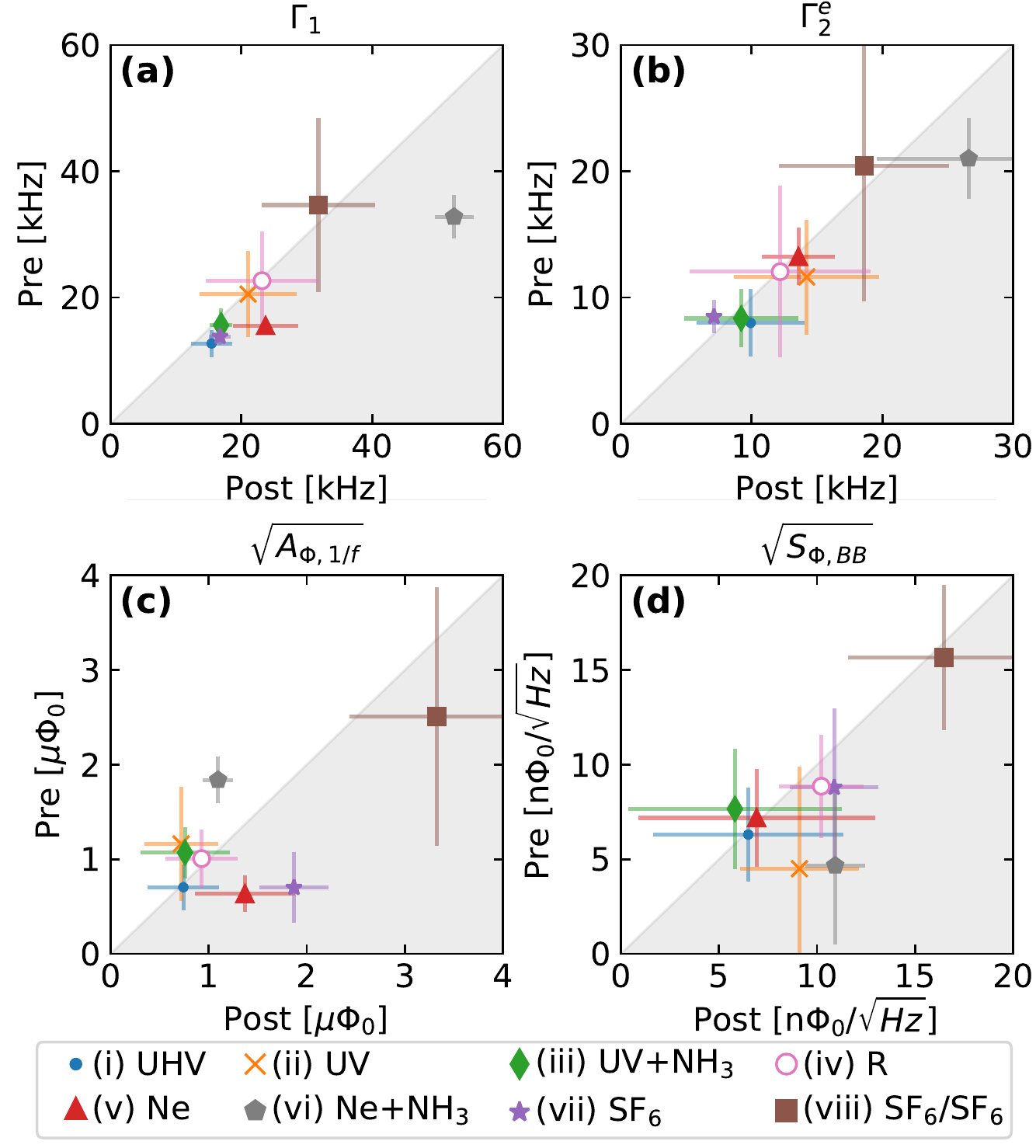}
	\caption{Impact of surface treatments listed in Tab.~\ref{tab:tab1} on flux noise and coherence parameters.
	Mean qubit parameters pre- (y-axis) and post-treatment (x-axis) are compared, where the grey shaded area indicates a degradation and conversely the white area of the graph an improvement in the parameters.
	(a) Weighted mean of \Gone~over the whole flux interval, with the standard deviation of the spread used as weights.
	Error bars indicate the weighted mean of the spread.
	(b) Mean value of \Gtwo~measured at the SSP of the qubits.
	Error bars indicate the standard deviation of values measured across multiple qubits.
	(c-d) Weighted mean values of \Af~(c) and \Sw~(d) extracted from quadratic fits to \Gtwophi~and weighted by the fit error.
	Error bars indicate the weighted standard deviation of the mean values.
	}
	\label{fig:fig3}
\end{figure}

We now compare the four coherence parameters \Gone=1/\Tone~ averaged over the entire flux interval, \Gtwo~$=1/$\Ttwoecho~at the SSP, $\sqrt{A_{\Phi,1/f}}$ and $\sqrt{S_{\Phi,\textrm{BB}}}$ between treatments as described in Tab.~\ref{tab:tab1}.
For each comparison we measured and analysed data of 2 - 22 qubits.
This is critical in order to disentangle effects of the treatments from the influence of cooldown-to-cooldown variations, which can be significant in systems like ours~\cite{Schlor2019}.
In Fig.~\ref{fig:fig3} we report weighted mean values of the four coherence parameters for each  surface treatment comparison.
The effect of the treatment on \Gone~averaged over the whole flux interval and \Gtwo~at the SSP is shown in panels (a) and (b) of Fig.~\ref{fig:fig3} respectively.
Since \Gone~is largely independent of flux and \Gtwo~is to first order insensitive to flux noise at the SSP, the observed effects of the treatments on these two parameters also apply to FFQs.
Panels (c) and (d) in Fig.~\ref{fig:fig3} present the effect of the surface treatments on the flux noise parameters \Af~and \Sw~, as defined above.

\setlength{\parindent}{0pt}
\paragraph*{(i) UHV:} This data compares coherence parameters obtained using standard \emph{ambient} packaging with those obtained in the UHV package.
This allows to assess if a better vacuum environment improves coherence parameters of the qubits and if the UHV package has a detrimental effect on the measured coherence values.
The data for UHV packaging in Fig.~\ref{fig:fig3} (filled circle) shows a slight increase in \Gone~(a) and \Gtwo~(b).
However, the \onef~(c) and broadband (d) flux-noise parameters do not seem to be effected by the packaging.
We believe that the small increase in \Gone~may be attributed to the ageing of the qubits, as this was the first comparison after qubit fabrication.
Furthermore, the first anneal of the qubit chip to $80^\circ$C in the treatment system may also have influenced this.
Our conclusion remains that there is no significant impact of the UHV package on these four measured coherence parameters.
The UHV package and packaging environment are thus suitable for comparing surface treatments.
Furthermore, it seems that ultra-high vacuum alone is not enough to significantly improve the coherence figures of the qubits.

\paragraph*{(ii) UV:} UV-light is frequently used as a gentle surface cleaning agent as it leaves solid surface materials such as oxides intact, while at the same time removing physiosorbed molecular species such as H$_2$O, H$_2$, O$_2$ or even some organic molecular residues.
This is expected to improve flux noise and specifically the slow \onef~type fluctuators~\cite{Kumar2016,18GraafAA,16GraafAA}.
Work on SQUIDs~\cite{Kumar2016} suggests an improvement of the \onef~noise for SiN$_x$ encapsulated devices but not for devices with SiO$_x$ which the authors attribute to the more efficient desorption of O$_2$ from the SiN$_x$ surface with UV light. However, in our experiments the UV light source has $2\times$ higher energy and roughly $100\times$ the intensity.
Our results show that UV light has no impact on \Gone, however it clearly improves the measured \onef~amplitude (see Fig.~\ref{fig:fig3}~(a,c) (x-symbols)).
\Gtwo~increases slightly, and broadband flux noise is increased by almost a factor two as shown in Fig.~\ref{fig:fig3}~(b,d).
The improvement of \onef~flux-noise by UV light can be expected from the removal of physiosorbed molecular species.
It may even be attributed to desorption of O$_2$ which has been indicated as a source of \onef~flux noise in Ref.~\cite{Kumar2016}.

The increase in broadband flux-noise and the slightly larger \Gtwo~indicate that there are added noise sources which are active at intermediate frequencies.
This should be in a range clearly below the qubit frequency f$_{01}$ since \Gone~is not affected by it.
The increase of \Gtwo~ at the flux insensitive SSP may indicate that the noise source is not entirely related to flux noise but also induces frequency noise mediated e.g. by near resonant two-level fluctuators~\cite{Schlor2019,15FaoroAA}.
We tentatively associate such noise with the creation of surface dangling bonds due to the high energy of the UV light ($\lambda\sim 140$~nm).
These may lead to charged (or magnetic) fluctuators either directly or by bonding e.g. with residual atomic O and H.

\paragraph*{(iii) \NH:} In order to examine whether the created dangling bonds can be passivated, we use a back-fill with non-magnetic \NH~which is known to be highly reactive and replace oxygen species due to the higher free energy of adsorption of \NH~compared to O$_2$~\cite{Kumar2016}.
Hence, we expect a suppression of magnetic flux noise, as previously shown for SQUIDs in Ref.~\cite{Kumar2016}. We compare the previous UV treatment with an identical one where additionally the package is back-filled with \NH~before transfer to the cryostat.
 From Fig.~\ref{fig:fig3}~(a-b) (diamonds) it is apparent that there is no significant change from the \NH~back-fill on \Gone~and \Gtwo.
 However, a clear improvement is found for the \onef~and the broadband flux-noise component, see Fig.~\ref{fig:fig3}~(c-d) respectively. This indicates not only that \NH~passivates the created dangling bonds but also further replaces magnetic molecules, such as O$_2$, on the chip surface (additional improvement of \onef-noise).
The neutral effect on \Gtwo~at the SSP means that some non-flux noise related dephasing remains after the UV treatment even with the passivation.
Both effects are in line with the observations in Ref.~\cite{Kumar2016}, where a reduction of the noise PSD in DC SQUIDs is achieved through \NH~passivation.

\paragraph*{(iv) Reference:}
Effects of UV-light exposure are limited to the topmost surface layers of the quantum chips and this process is expected to be fully reverted when a chip is exposed to ambient for extended periods of time.
Hence, we compare subsequent identical UV-treatments as a reference baseline.
The data in Fig.~\ref{fig:fig3}~(a-d) (unfilled circle) shows no significant change for any of the four coherence parameters.
This highlights the reliability of our method and indicates that UV-light exposure is a fully reversible treatment which we will include as a surface cleaning step for all treatments from here onwards.

\paragraph*{(v) Ne:} In contrast to UV light, ion milling also removes solid surface layers such as resist residues or thin oxides.
Such ion mill cleans with inert ions (Ar$^-$, Ne$^-$) are an integral part of Josephson junction fabrication for superconducting qubits, especially if the junctions are fabricated in a multi-stage process~\cite{17WuAA}.
However, ion milling has also been identified as a source of increased loss due to the created surface damage~\cite{Dunsworth2017, Nersisyan2019}.
In our case, we performed in-situ ion milling with Ne$^-$ ions at an angle of $\sim 30^\circ$ from the surface~\cite{Mergenthaler2021} and at low energy, in order to sputter-clean the chip surface just before closing the sample package. We compare this to a treatment with only UV-light irradiation.

The data in Fig.~\ref{fig:fig3}~(a) (triangle) shows a significant increase of \Gone~by about 50\%. Interestingly, \Gtwo~is very similar as in the treatment solely with UV-light, see Fig.~\ref{fig:fig3}~(b).
\Af~is increased and \Sw~is not affected by the Ne treatment, cf. Fig.~\ref{fig:fig3}~(c-d).
The degradation of \Gone~is most likely linked to the fact that ions penetrate the device materials more deeply and leave behind amorphous defects and vacancies also in the close vicinity of the junction.
This is especially true for our ion milling configuration, where the ions impinge on the sample chip under a shallow angle~\cite{Mergenthaler2021}.
Hence, neon ions can also hit the sidewalls of the Josephson junctions and may penetrate the junction oxide at the edge, remove oxide and create defects that couple more strongly to the qubit.
Simple simulations have shown that for our ion milling conditions Ne ions penetrate the surface to a depth of $2-5$~nm~\cite{Mergenthaler2021}, where defects and voids may be created.
Additionally, we expect that almost a nanometer of oxide is removed in the process.
The constant \Gtwo~and \Sw~seem to indicate that a similar amount of surface dangling bonds and related defects are created by neon ion milling as with UV irradiation.
However, the observed increase in \Af~shows that there are more slow magnetic fluctuators that remain on the chip surface after ion milling.
Considering that ion milling only partially sputters the native silicon oxide and junction oxides, it seems reasonable that adsorbed oxygen from redeposition or incomplete desorption may account for this.

Even for these relatively mild ion milling parameters there is a clear detrimental effect on \Gone~and we conclude that ion milling of the active qubit area (before and after qubit fabrication) with Ne or Ar should indeed be avoided if possible.

\paragraph*{(vi) Ne + \NH:} As a comparison to the previous treatment we perform a much more intense ion milling treatment with a 50~\% higher neon ion energy and double the milling duration. With these parameters we expect the native surface oxides to be completely sputtered away and ion damage to reach deeper into the surface layers~\cite{Mergenthaler2021}. However, we also use a \NH~backfill to passivate the surface after ion milling.
It is interesting that even though we expect much more ion damage with these parameters we find that \Gone~is only increased by $\sim 40$~\% compared to the UV baseline, see Fig.~\ref{fig:fig3}~(a) (pentagon).
Moreover, \Af~is significantly decreased indicating that either \NH~efficiently reduces adsorbed magnetic surface species as for the UV treatment or there is simply less residual oxygen from redeposition after fully removing the native surface oxides.
However, we find that both \Gtwo~and \Sw~increase significantly which is in line with the higher ion energy creating more and deeper defects, voids and unsaturated bonds that cannot be passivated with \NH~from the surface.

Therefore, \NH~passivation is again helpful but cannot fully overcome the detrimental effects of the high energy ions, especially below the surface.

\paragraph*{(vii) \sfsix:} With Ne gas our ion milling system accelerates individual atomic ions which are small and inert towards the chip surface. Thus, the ions easily penetrate into the top layers of the device.
When using \sfsix~gas in the same sputter system, larger ionized molecular fragments are accelerated towards the chip and we don't expect them to penetrate below the surface and create subsurface defects.
\sfsix~plasmas are also commonly used for etching Si in the Bosch process which has been previously indicated as a path to improved quality factors of superconducting resonators and qubits~\cite{Sandberg2012, 15BrunoAA, Chu2016}.
Using similar ion energies as for neon we find that bombardment of the chip surface with \sfsix~ions does not have any detrimental effect on \Gone~and \Gtwo, see Fig.~\ref{fig:fig3}~(a-b) (star).
This agrees with our expectation that fewer deep defects are created with \sfsix.
In addition, \sfsix~ions and radicals are known to readily react with silicon surfaces and either etch or passivate (depending on ion species and surface material)~\cite{81dAgostinoAA}.
We don't know if the native oxides of the silicon chip surface is fully removed with the ion mill conditions that we used here but we expect that some residual SiO$_x$F$_y$ species~\cite{95LegtenbergAA} or similar may remain on the chip surface.
From Fig.~\ref{fig:fig3}~(c-d) (star) we see that both flux-noise parameters show an increase and especially the \onef~noise component is significantly higher.
This seems to indicate that some of the residual surface molecules from the \sfsix~etch add to the population of fluctuating surface spins but are not electrically active i.e. these fluctuators would not impact the coherence of FFQs.

\paragraph*{(viii) \sfsix / \sfsix:} From consecutive treatments with \sfsix~[see Fig.~\ref{fig:fig3}~(a-b) (square)] we find that even though \Gone, \Gtwo~and \Sw~remain constant after the second \sfsix~treatment, the negative impact of \sfsix~residues on \Af~seems to be additive.
Interestingly, this means that the "passivation" layer containing residual surface spins is robust both in ambient and under UV irradiation in UHV.

In summary, the above treatment comparisons indicate that deep ion damage from neon sputtering leads to a reduction in \Tone~which cannot be easily mitigated with a surface treatment such as \NH.
Surface dangling bonds or subsequent reaction of these with adsorbed surface species can increase both electrical and flux-noise in an intermediate frequency range.
These noise sources can be partially passivated with \NH.
Low frequency \onef~flux-noise as measured by \Af~seems to be linked to adsorbed surface species that can usually be photo-desorbed with UV light.
This statement is, however, an extrapolation since \Ttwoecho~measurements only probe frequencies above a few 10~kHz.
We therefore aim to substantiate these results using long term noise spectroscopy to study the low-frequency noise limit of the qubits after each treatment.

\setlength{\parindent}{15pt}

\section{Low-Frequency Parameter Fluctuations}
\label{noise-spectroscopy}

\begin{figure}[t]
	\begin{center}
		\includegraphics[width=0.95\columnwidth]{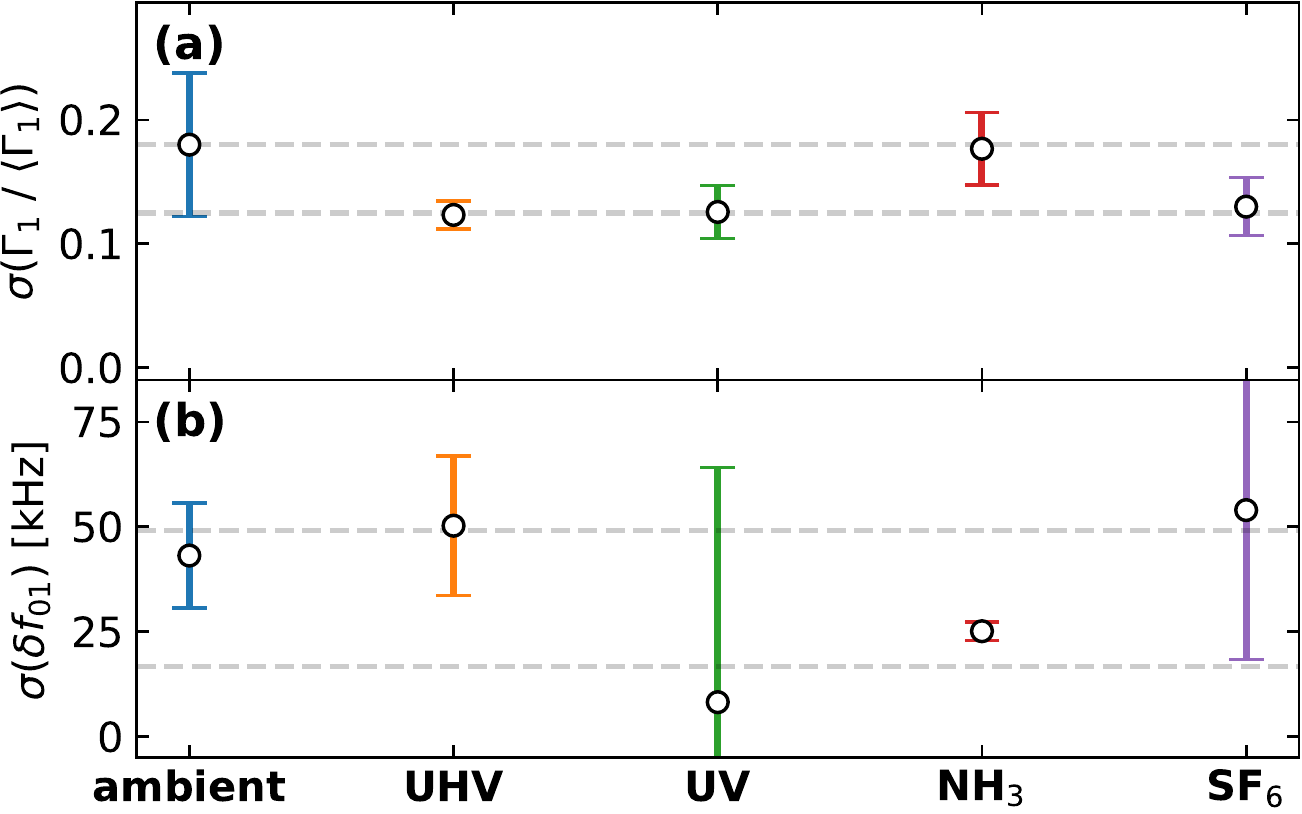}
	\caption{Width of the median distribution of the relative fluctuations in $\Gamma_{1} / \langle\Gamma_{1}\rangle$ (a) and
		of the frequency fluctuations $\delta f_{01}$ (b).
		Error bars denote $\pm$ one median absolute deviation of the distribution of widths. Dashed lines are guides to the eye.}
	\label{fig:fig4}
	\end{center}
\end{figure}

We use fast interleaved measurements of $\Gamma_{1}$ and $f_{01}$, calculated from Ramsey traces, to track low-frequency fluctuations of these parameters (see supplementary materials~\cite{supplement} for details).
In total each measurement consists of 10~h of continuous data acquisition, interspersed with short calibrations every 2.5~h to account for long-term parameter drifts.
Due to the long measurement duration we were forced to reduce the number of investigated treatments and qubits.
Furthermore, we did not perform a full comparative study as in Section~\ref{treatment-comparisons}, but instead we took just one dataset at the SSP for a selection of qubits.
We use integrated histograms of $\Gamma_{1}$ and $f_{01}$ to extract a median distribution width that allows us to quantify the low-frequency fluctuations of these parameters.
Fig.~\ref{fig:fig4} shows these noise spectroscopy results for the treatments indicated on the x-axis.

Panel (a) of Fig.~\ref{fig:fig4} shows the width of the $\Gamma_{1}$-distribution normalised with the mean $\langle\Gamma_{1}\rangle$ for each qubit separately.
We obtain the width of the distribution as the interval over which the normalized integral of the median over all histograms for a specific treatment changes from a value of 25\% to 75\%.
This corresponds to a width that contains half of all the histogram counts.
We use median histograms to reduce the impact of outliers in the data.
The error bars indicate the median absolute deviation of the widths of the integrated histograms for each qubit.
An example with a full dataset is given in the supplement~\cite{supplement}.

In the data of panel (a) we observe two different widths of the $\Gamma_{1}$-distributions indicated by the dashed lines as a guide to the eye.
Ambient exposure and a back-fill with \NH~lead to more fluctuations with a fluctuation-width of roughly 18\% of the average $\Gamma_{1}$.
On the other hand UHV, UV and \sfsix~ treatments show a reduced fluctuation width of around 13\% of the average $\Gamma_{1}$.
Here, we note that error bars for the ambient treatment are quite large and have to be considered when drawing conclusions for this treatment.
Since the data was measured at the SSP we don't expect flux noise to be responsible for the observed trends.
However, the low-frequency fluctuations on the SSP seem to be related to an increase in adsorbed surface species which is the case for both ambient and \NH~treatments.
These adsorbates contribute to an ensemble of low-frequency thermal fluctuators who's incoherent thermal dynamics affect $\Gamma_{1}$ indirectly through their coupling with high-frequency TLSs that in turn affect the qubit through changing dispersive shifts~\cite{Muller2015}.

Fig.~\ref{fig:fig4} panel (b) shows the distribution widths of frequency fluctuations $\delta f_{01}$ from noise spectroscopy, with widths and errorbars as defined above.
A first observation is that UV exposure appears to have a small positive effect on the frequency stability for the qubits, as both UV and \NH~datasets have been treated with UV (see dashed lines as a guide to the eye).
The backfill with \NH~does not appear to negatively affect the frequency fluctuations.
This seems to suggest that adsorbed species at the surface of the chip negatively impact the low-frequency electromagnetic environment of the qubits and depending on wether they are charged or carry a spin affect different parameters. A UHV environment may be enough to remove some larger charged or polar molecules but is not enough to affect smaller spin carrying species such as H$_2$ or O$_2$.  These are removed by UV-light or a \NH ~treatment. This is consistent with the observations from above for \Af~which was found to be lower after both UV and \NH~treatements.
However, the latter seems to leave slow charged fluctuators that indirectly affect $\Gamma_{1}$.
An \sfsix ~treatment on the other hand leaves a more robust surface layer that is not removed during exposure to ambient or UV-light and does not affect $\Gamma_{1}$. However, there seems to be a magnetic fluctuator associated with one or several of the \sfsix ~residues that affects the qubit frequency directly even at the SSP. This is also reflected by the observation in Section~\ref{treatment-comparisons} that \sfsix~treatments lead to an increase of flux noise compared to a UV treatment alone.

\section{Qubit Frequency Tuning}
\label{frequency-tuning}
\begin{figure}[t]
	\centering
	\includegraphics[width=0.95\columnwidth]{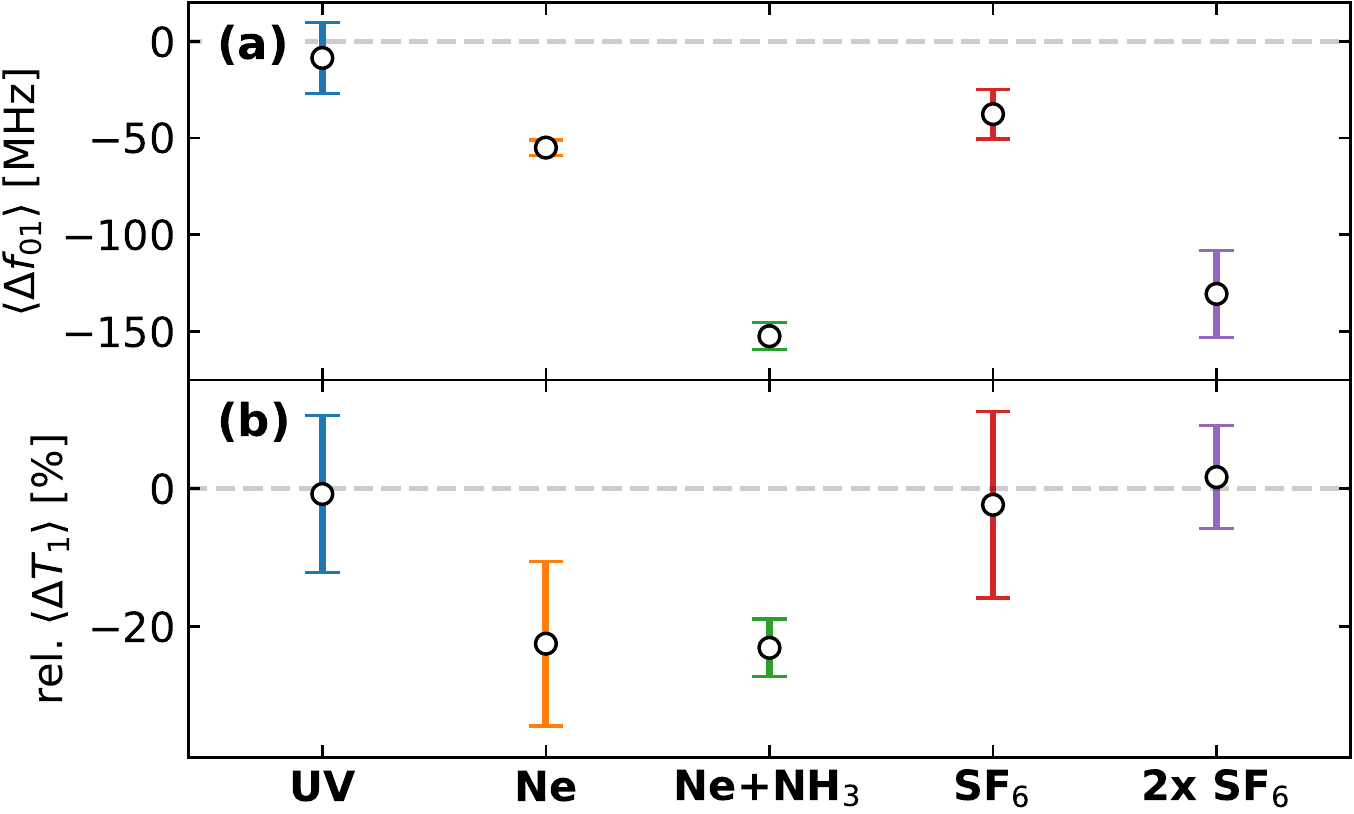}
	\caption{Changes in qubit frequency (a) and \Tone ~(b) at the SSP after ion milling treatments outlined in Section~\ref{treatment-descriptions}.
	Circles are mean values and error bars indicate standard deviation.}
	\label{fig:fig5}
\end{figure}

Beyond these coherence considerations, which are the main focus of the paper, it is interesting to see how other properties of the qubits such as $\langle T_{1}\rangle$ and $\langle f_{01}\rangle$ are affected on average by the treatments.
This is particularly true after ion milling treatments which remove material from the chip surface.
Fig.~\ref{fig:fig5} shows the average qubit frequency change $\langle \Delta f_{01}\rangle$ (a) and the relative change in $\langle T_1\rangle$ (b) at the SSP for the ion milling treatments as well as the UV treatment for comparison.
The UV treatment does not show any significant change in either $\langle f_{01}\rangle$ or $\langle T_{1}\rangle$.
We have already shown above that $\langle T_{1}\rangle$ is reduced by Ne ion-milling but not affected by ion milling with \sfsix.
We find, however, that the qubit frequency is reduced by all ion milling treatments.
This can be understood as a trimming of the two involved Josephson junctions.
Due to the shallow angle of the ion beam the treatments sputter material on the sidewalls of the junctions.
This leads to a small reduction in junction area which increases junction resistance, decreases the effective Josephson energy of the qubits and thus reduces the qubit frequency~\cite{Mergenthaler2021, Ambegaokar1963a}.
For treatments with higher ion energy or higher ion current the frequency change is larger, cf. Fig.~\ref{fig:fig5}~(a) Ne vs Ne+\NH.
This is expected because more material is removed and the junction shrinks accordingly.
A confirmation of this comes from measurements of the junction resistance in junction arrays fabricated simultaneously with the qubit chips (see supplementary materials for more details~\cite{supplement}).
We find that for Ne the junction resistance empirically changes as $\frac{\Delta R}{R} \approx 0.5~\mathrm{U_{acc}[V]}~\mathrm{t[s]}~\mathrm{I_{ion}[A]}$ with $R\approx29~\mathrm{k\Omega}$ at the outset for each of the two junctions in the SQUID loop. %% This was dR[Ohm] = 0.05 U[V]t[h]I[nA]
For long treatments with high ion energy (e.g.~Ne+\NH) we see a reduction of this rate which is due to the lower sputter yield of Al compared to e.g. Al$_2$O$_3$ and SiO$_2$ when the surface oxide layers are fully removed~\cite{Mergenthaler2021}.
Furthermore, if we use the \emph{Ambegaokar-Baratoff} relation~\cite{Ambegaokar1963a} to determine the change in qubit frequency we find $\langle\Delta f_{01}\rangle = -\frac{1}{2 h} \sqrt{\Delta\cdot E_C}\sqrt{\frac{R_K}{R_N}}\frac{\Delta R_N}{R_N}\approx-1.866~\mathrm{GHz}\sqrt{\frac{25.8k\Omega}{R_N}}*\frac{\Delta R}{R_N}$ with $R_N=R/2$. This is consistent over all neon treatments (see supplementary materials ~\cite{supplement}).

For \sfsix~we did not perform a detailed study of the junction resistance but it seems that for the same ion energy and ion flux the measured increase in resistance $\Delta R$ is larger than for neon.
In contrast to this, the change in frequency is smaller for \sfsix~ion milling than with neon.
This may be due to  the passivation layer that is left after \sfsix~treatments.
Such a passivation could both prevent re-oxidation of the junction after the treatment (less reduction in junction area) and reduce surface conductivity in the ambient junction resistance measurements (higher apparent resistance of the junctions).

An important conclusion is that ion milling with both ion species allows to trimm the qubit frequency easily by more than 100~MHz.
At the same time we do not see any detrimental effects on $\langle T_{1}\rangle$ after the \sfsix~treatment while neon yields a decrease in $\langle T_{1}\rangle$ by 20\% or more if $\langle f_{01}\rangle$ is trimmed (c.f. Fig.~\ref{fig:fig5}~(b)).
Considering that flux-noise increases after \sfsix~treatments as shown above, this still makes \sfsix~ion milling an interesting method to trimm the qubit frequency of fixed frequency transmon qubits which are insensitive to flux-noise.
Such a capability is critical due to the strict margins for frequency detuning between qubit pairs in corresponding quantum computing systems~\cite{Hertzberg2020}.
To date, post fabrication frequency trimming of fixed frequency transmon qubits has been demonstrated with local laser annealing of the Josephson junction~\cite{Hertzberg2020}, which also reduces the qubit frequency but may also deposit residues on the chip surfaces.
Our method can easily be extended to use a focused ion beam and would not only allow for local tuning but may also be beneficial in terms of cleaning the Josephson junction environment from residues and contamination~\cite{Mergenthaler2021}.

\section{Conclusion}
\label{conclusion}
 In conclusion, under UHV we have studied a variety of surface treatments and their effect on coherence and noise parameters of flux-tunable transmon qubits.
 Except for neon ion milling, none of the treatments lead to strong changes in either \Gone~or \Gtwo~ at the SSP.
 In contrast, the flux noise parameters \Af~and \Sw~are improved when removing magnetic adsorbates from the surfaces, i.e. after UV light exposure as well as after UV light exposure and consecutive \NH~surface passivation.
 Effects from UV light exposure are reset when exposing the qubits to ambient atmosphere while those of \sfsix ~treatments seem to be additive.
 We also investigated slow qubit parameter fluctuations with long-term noise spectroscopy.
 Here, our results suggest that adsorbed species have a negative effect on fluctuations in \Gone.
 Similarly, fluctuations of the qubit frequency seem to be reduced by removing adsorbates from the chips surfaces with UV light.

 Average qubit frequency $\langle f_{01}\rangle$ and $\langle T_{1}\rangle$ of our qubits were mostly affected by the more invasive ion milling treatments. We find that both ion milling with Ne ions and \sfsix ~ions can be used to selectively trimm the qubit frequency post fabrication by more than 100~MHz.
 Whilst Ne ions have a detrimental effect on $\langle T_{1}\rangle$, \sfsix ~ions on average do not degrade $\langle T_{1}\rangle$ of the qubits.
Hence, the latter is suitable to frequency trimm FFQs, which can be beneficial for the development of larger scale quantum processors based on transmon qubits that have tight qubit frequency margins.

\acknowledgments{We thank I. Tavernelli for insightful discussions and R. Heller, H. Steinauer, A. Zulji and S. Gamper for technical support. A.F., P.M. and S.F. acknowledge support form IARPA LogiQ program under contract W911NF-16-1-0114-FE for design and characterization of the UHV package. A.F., G.S. and C.M. acknowledge support from the Swiss National Science Foundation through NCCR QSIT.}

%\bibliography{CoherencePackaging}
%\bibliography{library}


\end{document}